\documentclass[10pt,prl,superscriptaddress,preprintnumbers,balancelastpage,twocolumn]{revtex4-2}
\usepackage{graphicx}
\usepackage{epstopdf}
\usepackage{bm}
\usepackage[utf8]{inputenc}
\usepackage[hidelinks,colorlinks]{hyperref}
\usepackage{color}
\usepackage{soul}
\usepackage{amsmath}
\usepackage{amssymb}
\usepackage{mathtools}
\usepackage{xfrac}
\usepackage{rotating}
\usepackage{enumitem}
\usepackage{scrextend}
\usepackage{slashed}
\usepackage{subfig}
\usepackage{braket}

\newcommand{\al}{\alpha}

\newcommand{\mm}{\mathcal{M}}
\newcommand{\eps}{\epsilon}
\newcommand{\tts}{\mathbf{T}_t^2}
\newcommand{\tsu}{\mathbf{T}_{s-u}^2}

\newcommand{\mTree}{\mathcal{M}^{\text{tree}}}

\newcommand{\mExpM}[2]{\hat{\mathcal{M}}^{(-,#1,#2)}}
\newcommand{\tij}{\mathbf{T}_i\cdot\mathbf{T}_j}

\begin{document}

\title{Climbing three-Reggeon ladders: 
four-loop amplitudes \\ in the high-energy limit in full colour}

\author{Giulio Falcioni}
\email{Giulio.Falcioni@ed.ac.uk}
\affiliation{Higgs Centre for Theoretical Physics, School of Physics and Astronomy, The University of Edinburgh, Edinburgh EH9 3FD, Scotland, UK}

\author{Einan Gardi}
\email{Einan.Gardi@ed.ac.uk}
\affiliation{Higgs Centre for Theoretical Physics, School of Physics and Astronomy, The University of Edinburgh, Edinburgh EH9 3FD, Scotland, UK}

\author{Calum Milloy}
\email{CalumWilliam.Milloy@unito.it}
\affiliation{Dipartimento di Fisica and Arnold-Regge Center, Universit\'{a} di Torino,
  and INFN, Sezione di Torino, Via P. Giuria 1, I-10125 Torino, Italy}

\author{Leonardo Vernazza}
\email{Leonardo.Vernazza@to.infn.it}
\affiliation{Dipartimento di Fisica and Arnold-Regge Center, Universit\'{a} di Torino,
and INFN, Sezione di Torino, Via P. Giuria 1, I-10125 Torino, Italy}
\date{\today}

\begin{abstract}
Using an iterative solution of rapidity evolution equations, we compute partonic $2\to 2$ gauge theory amplitudes at four loops in full colour up to the Next-to-Next-to-Leading Logarithms (NNLL) in the Regge limit. 
By contrasting the resulting amplitude with the exponentiation properties of soft singularities we determine the four-loop correction to the soft anomalous dimension at this logarithmic accuracy, which universally holds in any gauge theory.  We find that the latter features quartic Casimir contributions beyond those appearing in the cusp anomalous dimension. Finally, in the case of ${\cal N}=4$ super Yang-Mills, we also determine the finite hard function at four loops through NNLL in full colour.
\end{abstract}
\maketitle

The high-energy limit of gauge-theory scattering amplitudes 
has long been a source of unique insight into gauge 
dynamics.  Amplitudes drastically simplify in this limit, and their factorisation in
rapidity reveals new degrees of freedom that propagate in two transverse dimensions.
Rapidity evolution equations, BFKL \cite{Kuraev:1977fs,Balitsky:1978ic} and its non-linear generalisation~\cite{Balitsky:1995ub,*Kovchegov:1999yj,*JalilianMarian:1996xn,*JalilianMarian:1997gr,*Iancu:2001ad},  manifest concepts from Regge theory~\cite{Collins:1977jy}, leading to remarkable new insights.
Several avenues have emerged in recent years which exploit the predictive power of factorization and evolution in
rapidity, and translate it into concrete predictions for partonic amplitudes \cite{Dixon:2012yy,Dixon:2014voa,Caron-Huot:2013fea,Caron-Huot:2017fxr,Caron-Huot:2017zfo,Caron-Huot:2020grv,Gardi:2019pmk,DelDuca:2019tur,Bartels:2020twc}. New results regarding the Regge limit have been instrumental in determining multi-leg planar $ \mathcal{N} $ = 4 super Yang-Mills amplitudes in general kinematics to unprecedented accuracy, see e.g.  \cite{Dixon:2014voa,Caron-Huot:2016owq,Caron-Huot:2019vjl}. 
In parallel, computation of the Regge limit in $2\to 2$ scattering for general colour~\cite{Caron-Huot:2013fea,Caron-Huot:2017fxr,Caron-Huot:2017zfo,Caron-Huot:2020grv} have been shown to provide powerful constraints on soft singularities of amplitudes in general kinematics~\cite{DelDuca:2011ae,Henn:2016jdu,Almelid:2017qju,Gardi:2019pmk}.

Our focus here is on universal features of  $2\to 2$ gauge-theory scattering amplitudes, $1+2\to 3+4$, in the high-energy limit, following~\cite{Caron-Huot:2013fea,Caron-Huot:2017fxr,Caron-Huot:2017zfo,Caron-Huot:2020grv,Gardi:2019pmk}. These amplitudes are described by two independent Mandelstam invariants $s\equiv (p_1+p_2)^2$ and $t\equiv  (p_1-p_4)^2$, with $u\equiv  (p_1-p_3)^2=-s-t$, where in the high-energy limit $s\gg -t$, and we apply perturbation theory, assuming that the momentum transfer $-t$ is large compared to the QCD scale.
For $s\gg -t$ the perturbative amplitude is dominated by large logarithms in the ratio $\frac{s}{-t}$. Famously, the Leading Logarithms (LL) can be resummed to all orders~\cite{Lipatov:1976zz,Kuraev:1976ge} via
\begin{equation}
\mm^{\text{LL}}_{ij\to ij}(s,t) 
=  \left(\frac{s}{-t}\right)^{C_A \,\al_g(t)}\!{\cal M}^{\rm tree}_{ij\to ij},
\label{eq:ReggeTraj}
\end{equation}
where ${\cal M}^{\rm tree}_{ij\to ij}= g_s^2\frac{2s}{t} \tij$ is the tree-level amplitude with the generator $\mathbf{T}_i$ in the representation of 
parton~$i$, and
\begin{equation}
\label{al_g}
\al_g
= \frac{\al_s}{\pi}\frac{r_\Gamma}{2\eps}+{\cal O}(\al_s^2)\,;\quad 
r_\Gamma = e^{\eps\gamma_E}\frac{\Gamma^2(1-\eps)\Gamma(1+\eps)}{\Gamma(1-2\eps)}
\end{equation} 
 is the gluon Regge trajectory, presented here in dimensional regularization with $D=4-2\epsilon$, where we suppressed higher-order corrections which contribute beyond~LL. The simple exponentiation property in~(\ref{eq:ReggeTraj}), with the characteristic colour charge $C_A$, 
 can be understood as due to the exchange of a single \emph{Reggeized gluon} (dubbed \emph{Reggeon}), which
admits a trivial evolution equation in rapidity.  At higher logarithmic accuracy more complex analytic structure 
emerges,  associated with compound states of multiple Reggeons~\cite{Low:1975sv,Nussinov:1975mw,Gunion:1976iy,Caron-Huot:2013fea}. 
The corresponding evolution equations are difficult to solve~\cite{Lipatov:1993yb,Faddeev:1994zg}, but they can be integrated iteratively~\cite{Caron-Huot:2013fea,Caron-Huot:2017fxr,Caron-Huot:2017zfo,Caron-Huot:2020grv}, to obtain perturbative high-energy amplitudes order-by-order in $\alpha_s$. In this paper we extend these methods to four loops and NNLL accuracy.

In the Regge limit amplitudes naturally split according to their \emph{signature} symmetry: \hbox{$\mm = \,\,\mm^{(-)} + \mm^{(+)}$} where $\mm^{(-)}$ and $\mm^{(+)}$ are respectively odd and even under $s\leftrightarrow u$. Upon using a signature-symmetric definition for the large logarithm,
\begin{equation}
\label{L_def}
 \frac12\left(\log\frac{-s-i0}{-t}+\log\frac{-u-i0}{-t}\right) = \log\left|\frac{s}{t}\right| -i\frac{\pi}{2}\equiv L \,,
\end{equation}
and expanding $\mm^{(-)}$ and $\mm^{(+)}$ according to
\begin{equation}
{\mm}^{(\pm)}_{ij\to ij} = \sum_{n=0}^\infty \left(\frac{\al_s}{\pi}\right)^n \sum_{m=0}^nL^m{\mm}^{(\pm,n,m)}_{ij\to ij},\label{eq:expansionDef}
\end{equation}
with ${\mm}^{(-,0,0)}_{ij\to ij}\equiv {\cal M}_{ij\to ij}^{\rm tree}$, it can be shown~\cite{Caron-Huot:2017fxr} that the odd amplitude coefficients $\mm^{(-,n,m)}_{ij\to ij}$ are purely \emph{real}, while the even ones $\mm^{(+,n,m)}_{ij\to ij}$ are purely \emph{imaginary}. 
Since Bose symmetry links the kinematic dependence to that of colour,  $\mm^{(+)}$ and $\mm^{(-)}$ are governed by $t$-channel exchange of colour representations which are respectively even and odd under $1\leftrightarrow 4$ (or $2\leftrightarrow 3$) interchange. The latter consist of an odd number of Reggeons, while the former an even number~\cite{Caron-Huot:2013fea}. Signature is preserved under rapidity evolution, and this greatly simplifies the computation of these amplitudes~\cite{Caron-Huot:2013fea,Caron-Huot:2017fxr,Caron-Huot:2017zfo,Caron-Huot:2020grv}.
The signature even amplitude $\mm^{(+)}$, which starts at NLL accuracy is governed at this logarithmic order by two-Reggeon exchange, satisfying BFKL evolution. 
Using an iterative solution of this equation, this entire tower of logarithms has recently been determined \cite{Caron-Huot:2013fea,Caron-Huot:2017zfo,Caron-Huot:2020grv}.

The present paper focuses on the signature odd amplitude. At NLL accuracy, $\mm^{(-)}_{ij\to ij}$ is still governed by a single Reggeon exchange (Regge pole)  \cite{Fadin:2006bj,*Fadin:2015zea}, with ${\cal O}(\alpha_s^2)$ corrections to the trajectory in (\ref{al_g}) \cite{Fadin:1995xg,*Fadin:1996tb,*Fadin:1995km,*Blumlein:1998ib} along with $s$-independent impact factors~\cite{Caron-Huot:2017fxr}:
\begin{equation}
\label{Regge-Pole-General}
\mm^{(-),\,\text{LL+NLL}}_{ij\to ij}  =
 e^{C_A \alpha_g(t) L} 
Z_i(t) D_i(t) \, Z_j(t) D_j(t)
{\cal M}^{\rm tree}_{ij\to ij}\,.
\end{equation}
The factors $Z_{i}(t)$ and $D_i(t)$ admit perturbative expansions with coefficients $Z^{(n)}_i$ and $D^{(n)}_i$ where the one-loop terms, $n=1$, contribute at NLL, while the higher-order terms give rise to further subleading logarithms. $Z_i(t)$ generate collinear singularities \cite{DelDuca:2011ae}
\begin{equation}
Z_i(t)=\exp\left\{-\frac{1}{2}\int_0^{\mu^2}\frac{d\lambda^2}{\lambda^2}\Gamma_i\left(\al_s(\lambda^2),\lambda^2\right)\right\},
\label{Zi}
\end{equation}
with $\Gamma_i = \frac{1}{2}\gamma_K\left(\alpha_s(\lambda^2)\right)C_i\log\frac{-t}{\lambda^2}+2\gamma_i$. Here $\gamma_K$ is the universal cusp anomalous dimension \cite{Korchemsky:1987wg,Gardi:2009qi,*Gardi:2009zv,*Becher:2009cu,Becher:2019avh}, $C_i$ is the quadratic Casimir in the representation of parton $i$ and $\gamma_i$ are anomalous dimensions associated with on-shell form factors~\cite{FormFactors,DelDuca:2014cya}. The collinear-subtracted impact factors $D_{i}(t)$ are known to two loops \cite{Caron-Huot:2017fxr}.

Our new computation concerns the NNLL tower, which manifests a Regge cut in the real part of the amplitude \cite{DelDuca:2001gu,DelDuca:2013ara,DelDuca:2014cya,Caron-Huot:2013fea,Fadin:2016wso,*Fadin:2017nka}. This phenomenon is associated to the exchange of three Reggeons, as shown by direct calculations to three loops \cite{Caron-Huot:2017fxr,Fadin:2016wso}. Here we take a further step by showing that the entire NNLL tower, ${\mm}^{(-,n,n-2)}$, can be computed using the evolution of one and three Reggeons and the transitions between them. We then explicitly compute the NNLL amplitude at four loops in full colour. Finally, upon comparing the result with the known exponentiation properties of infrared singularities we gain a powerful check and determine the soft anomalous dimension in the same approximation.

{\bf Methodology.} We describe $2\to 2$ scattering at high energy following \cite{Caron-Huot:2013fea}. Fast particles moving in the plus $(+)$ lightcone direction appear as infinite Wilson lines \cite{Korchemskaya:1994qp,*Korchemskaya:1996je}
\begin{align*}
e^{ig_s\mathbf{T}^aW^a(z)}\equiv
 \mathcal{P}\exp\!\left[ig_s \mathbf{T}^a\!\!\int_{-\infty}^{+\infty}\!\!\!dx^+ A^a_+(x^+,x^-\!=\!0,z) \right]
\end{align*}
at transverse position $x_{\perp}=z$,
where the generator $\mathbf{T}^a$ is in the representation of the scattering parton and $W$ identifies a Reggeon field. Rapidity divergences are regulated by introducing a cutoff $\eta=L$. The projectile and target in the scattering process, denoted respectively as $\ket{\psi_i}$ and~$\bra{\psi_j}$, are expanded in Reggeon fields, regulated at different rapidities. These are then evolved to equal rapidities by applying the Balitsky-JIMWLK Hamiltonian $H$ \cite{Balitsky:1995ub,*Kovchegov:1999yj,*JalilianMarian:1996xn,*JalilianMarian:1997gr,*Iancu:2001ad}. The contraction of Reggeons of equal rapidity is evaluated in terms of free propagators $\bra{W^a(p)}W^b(q)\rangle=\frac{i}{p^2}\delta^{ab}\delta^{2-2\epsilon}(p-q)+{\cal{O}}(g_s^2)$~\cite{Caron-Huot:2013fea}.  To compute the amplitude
\begin{equation}
\frac{i}{2s}\frac{1}{Z_iZ_j}\mm_{ij\to ij} = \braket{\psi_j|e^{-HL}|\psi_i}\, \label{eq:ampBraket}
\end{equation} 
we expand the projectile and target in the number of Reggeons $n$, such that $\ket{\psi_i} = \sum_{n=1}^\infty \left(r_\Gamma\alpha_s\right)^{(n-1)/2} \ket{i_{n}}$, where
\begin{align}
  \begin{split}
  \label{eq:ketexp}
 \ket{i_{n}} =\ket{i_{n}}^{\text{LO}}+\sum_{k=1}^{\infty}\left(\frac{\alpha_s}{\pi}\right)^k\ket{i_{n}}^{\text{N}^k\text{LO}}\,,
  \end{split}
\end{align}
with the tree-level normalized as $\braket{j_1|i_1}=\frac{ig_s^2}{t} \tij$.
The Hamiltonian $H$ allows transitions between states with different numbers of Reggeons \cite{Caron-Huot:2013fea}. However, signature symmetry  excludes transitions between states of odd and even numbers of Reggeons, e.g. $\braket{j_{3}|e^{-H L}|i_{2}} = 0$. Only odd (even) transitions contribute to the odd (even) amplitude. We denote individual Hamiltonians converting an~$n$ Reggeon state to a~$k$ Reggeon state as $H_{n\to k}$. Defining channels of colour 
flow~\cite{Dokshitzer:2005ek} by
\begin{align}
\mathbf{T}_s = \mathbf{T}_1 + \mathbf{T}_2, &\,& \mathbf{T}_t = \mathbf{T}_1 + \mathbf{T}_4,  &\,& \mathbf{T}_u = \mathbf{T}_1 + \mathbf{T}_3\,,
\end{align}
the single Reggeon evolves with \hbox{$H_{1\to 1} = -\tts\,\al_g(t)$}, 
with  $\bra{j_1} \tts  \ket{i_{1}}=C_A \braket{j_1 |i_{1}}$;
NNLL accuracy requires ${\cal O}(\alpha_s^3)$ corrections in $\al_g(t)$ of eq.~(\ref{al_g}). 

Multi-Reggeon transitions $H_{n\to k}$ were computed to leading order in \cite{Caron-Huot:2017fxr} by expanding the Balitsky-JIMWLK Hamiltonian~\cite{Balitsky:1995ub,*Kovchegov:1999yj,*JalilianMarian:1996xn,*JalilianMarian:1997gr,*Iancu:2001ad} in the Reggeon field, obtaining~\cite{Caron-Huot:2013fea,Caron-Huot:2017fxr}
\begin{align}
  \begin{split}
  \label{eq:Hexpand}
  H= \sum_{n\geq 1}\Big[H_{n\to n}^{\rm LO}+\sum_{l>0} (\alpha_s r_\Gamma)^l H_{n\to n\pm 2l}^{\rm LO}\Big]+\ldots\,,
\end{split}
\end{align}
where the ellipsis stand for ${\cal O}(\alpha_s^2)$  corrections to~$H$~\cite{Balitsky:2001mr,*Balitsky:2008zza,*Gardi:2006rp,*Balitsky:2013fea,*Kovner:2013ona}; these are not required for multi-Reggeon transitions at NNLL.
To determine ${\mm}^{(-,n,n-2)}$ it is convenient to extract $H_{1\to 1}$ in~(\ref{eq:ampBraket}), defining the \emph{reduced amplitude},
\begin{align}
 \frac{i}{2s}\hat{\mm}_{ij\to ij} \equiv\braket{\psi_j|e^{-\left(H - H_{1\to 1}\right)L}|\psi_i} =& \braket{\psi_j|e^{-\hat{H}X}|\psi_i},
\label{eq:reducedAmpBraket}
\end{align}
along with a \emph{reduced Hamiltonian} $\hat{H}$  and an expansion parameter $X\equiv \frac{\al_s}{\pi}{r_\Gamma}L$. Note that $\hat{H}_{1\to 1}=0$.
Expanding~(\ref{eq:reducedAmpBraket}) and collecting the NNLL in the odd amplitude we find, to all orders in $\al_s$,
\begin{align}
  \begin{split}
    \label{eq:mhatNNLL}
    &\frac{i}{2s}\hat{\mm}^{(-),\text{NNLL}}_{ij\to ij}\!=\left(\frac{\alpha_s}{\pi}\right)^2\Bigg\{r_{\Gamma}^2\pi^2 \Bigg[\sum_{k=0}^{\infty}\frac{(-X)^k}{k!}\braket{j_{3}|\hat{H}_{3\to3}^k|i_{3}}  \\ 
      &+\sum_{k=1}^{\infty}
\frac{(-X)^{k}}{k!}
\left[
\braket{j_{1}|\hat{H}_{3\to1}\hat{H}_{3\to3}^{k-1}|i_{3}}+
\braket{j_{3}|\hat{H}_{3\to3}^{k-1}\hat{H}_{1\to3}|i_{1}}\right]\\ 
      &+\sum_{k=2}^{\infty}\frac{(-X)^{k}}{k!}\braket{j_{1}|\hat{H}_{3\to1}\hat{H}_{3\to3}^{k-2}\hat{H}_{1\to3}|i_{1}}\Bigg]^{\text{LO}}\!\!\!\!+\braket{j_{1}|i_{1}}^{\text{NNLO}}\!\!\Bigg\}\,
  \end{split}    
\end{align}
With the sole exception of the term $\braket{j_{1}|i_{1}}^{\text{NNLO}}$, which is extracted from two-loop amplitudes~\cite{Caron-Huot:2017fxr}, eq.~(\ref{eq:mhatNNLL}) describes the whole NNLL tower using the leading-order formalism! 
\begin{figure}
  \subfloat[\label{fig:3to3Tree}]{\includegraphics[width=0.12\textwidth]{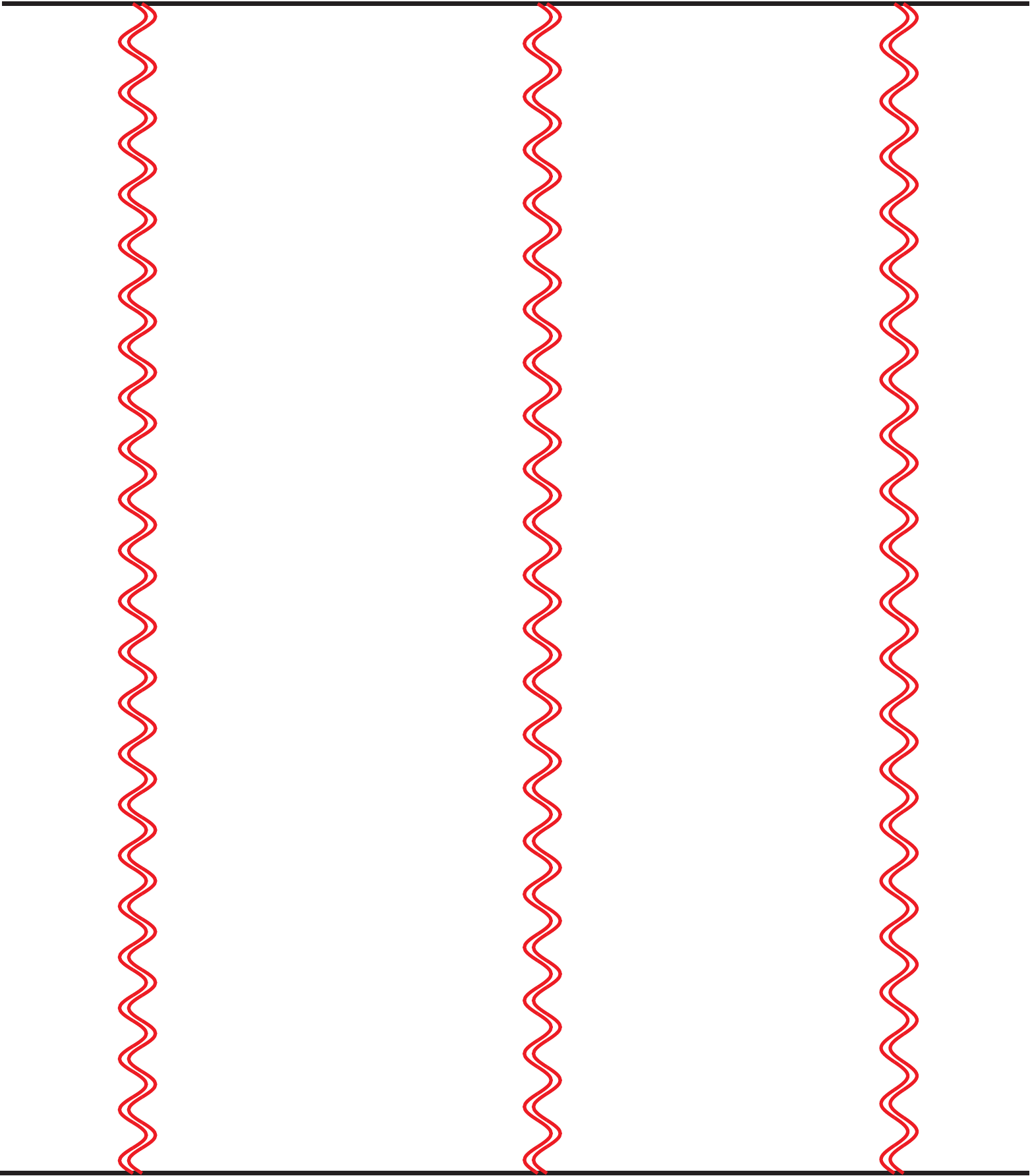}}\,\hspace{10pt}
  \subfloat[\label{fig:3to3Loop}]{\includegraphics[width=0.12\textwidth]{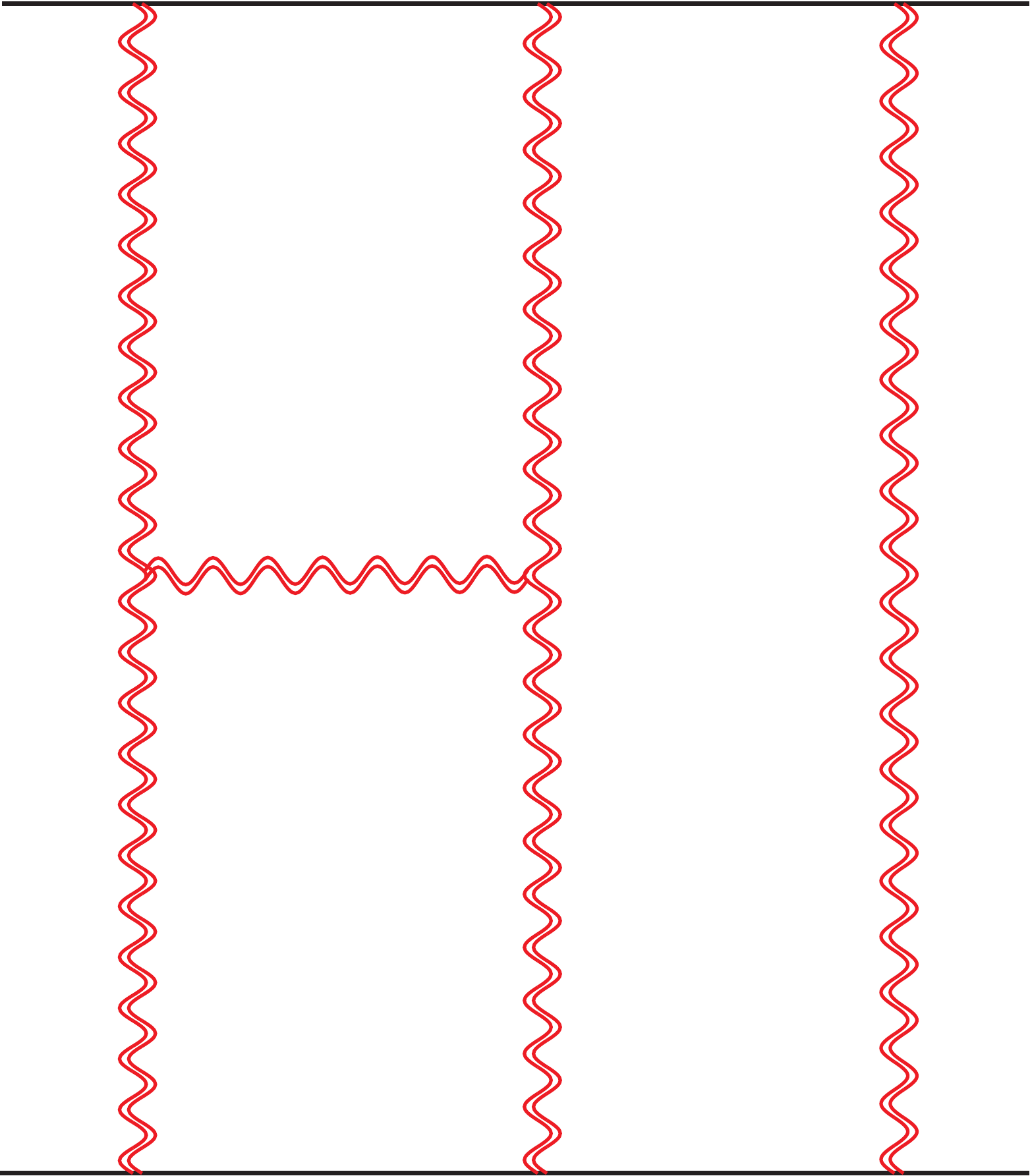}}\,\hspace{10pt}
  \subfloat[\label{fig:1to3}]{\includegraphics[width=0.12\textwidth]{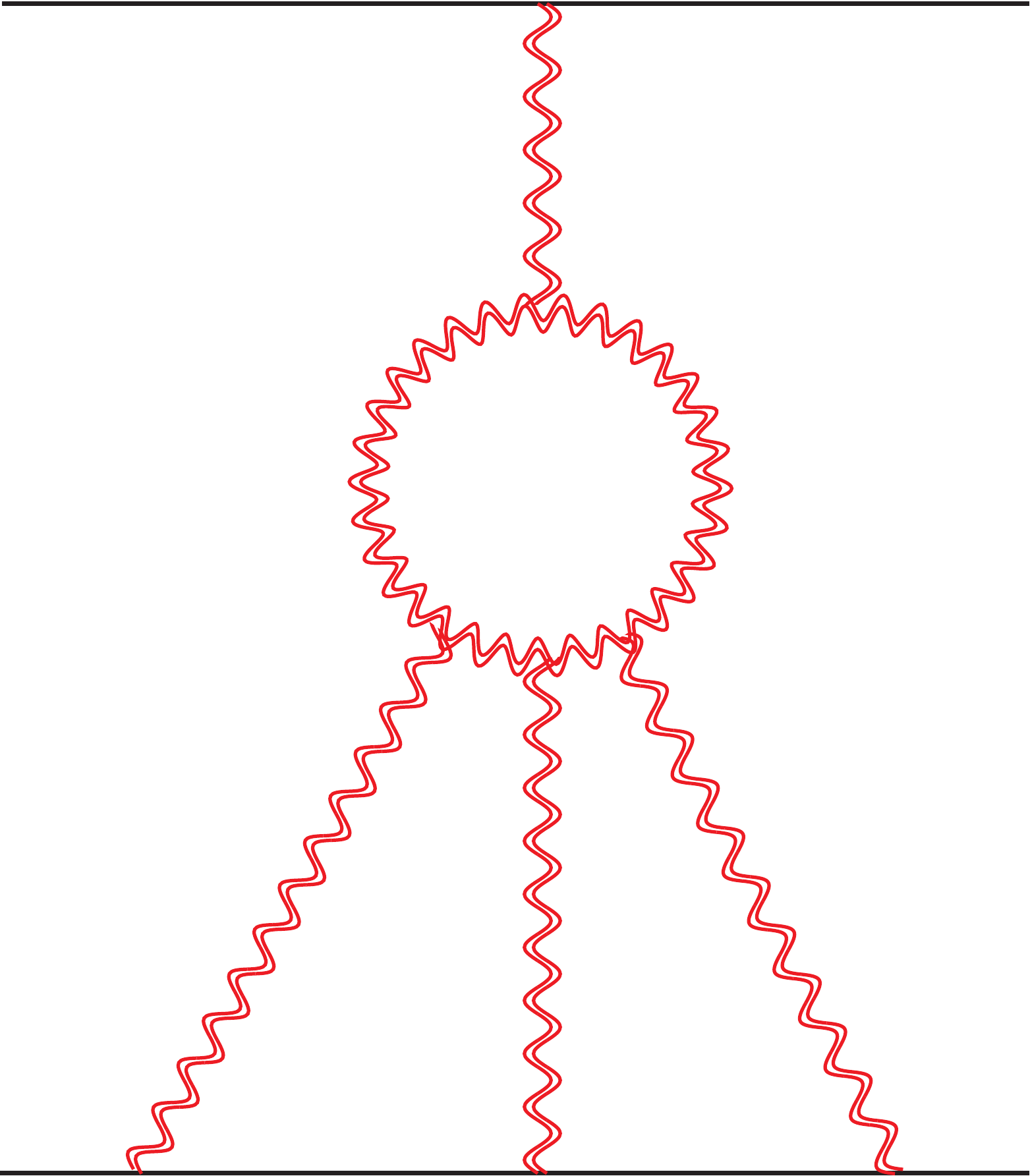}}\,
  \caption{Three-Reggeon exchange at two and three loops (a and b) and their mixing into a single Reggeon~(c).}
\end{figure}

{\bf Results at NNLL.} We proceed to evaluate $\hat{\mathcal{M}}$ in~(\ref{eq:mhatNNLL}) to four loops, following the notation of~(\ref{eq:expansionDef}).\\
{\it \textbf{Two loops.}} According to~(\ref{eq:mhatNNLL}) there are two distinct contributions to the two-loop amplitude $\mExpM{2}{0}$.
The first is the single Reggeon exchange, $\braket{j_1|i_{1}}^{\text{NNLO}}$, which may be read off eq.~(\ref{Regge-Pole-General}):
\begin{align}
\label{NNLL_single_Reggeon}
\braket{j_{1}|i_{1}}^{\text{NNLO}}=\left(D_i^{(2)}+D_j^{(2)}+D_i^{(1)}D_j^{(1)}\right)\braket{j_{1}|i_{1}}.
\end{align}
The second is the three Reggeon exchange depicted in Fig.~\ref{fig:3to3Tree} (where there is an implicit sum over permutations of the three-Reggeon attachment to the Wilson line):
\begin{eqnarray}
\label{33braket_two_loops}
&\braket{j_{3}|i_{3}} = -3\left(\frac{1}{\eps^2}-6\eps f_\eps \right) &\mathbf{C}_{33}^{(2)}\braket{j_{1}|i_{1}}\\
&f_\eps \equiv \zeta_3+\frac{3}{2}\eps\zeta_4+ \mathcal{O}(\eps^2)\,;
&\quad \mathbf{C}_{33}^{(2)}\equiv \frac{1}{24}\left((\tsu)^2-\frac{C_A^2}{12}\right)\!\nonumber
\end{eqnarray}
where $\tsu \equiv \frac{1}{2}(\mathbf{T}_s^2-\mathbf{T}_u^2)$. Adding up the two contributions
we obtain $\mExpM{2}{0}$ in agreement with \cite{Caron-Huot:2017fxr}.

{\it \textbf{Three loops.}} There are two contributions to $\mExpM{3}{1}$ in eq.~(\ref{eq:mhatNNLL}). The first, in Fig.~\ref{fig:3to3Loop}, is the $\hat{H}_{3\to 3}$ evolution,
\begin{eqnarray}
\label{eq:h333loops}
&\braket{j_{3}|\hat{H}_{3\to 3}|i_{3}} =
\Bigg[& \frac{1}{\eps^3}\left(C_A-\frac{5}{6}\tts \right)+ 2f_{\eps}\times \\ \nonumber
&&  \left(C_A-\frac{41}{6}\tts \right) +{\cal O}(\epsilon^2) \Bigg]\, \mathbf{C}_{33}^{(2)}\,\braket{j_{1}|i_{1}}\,,
\end{eqnarray}
which relates to $\mathbf{C}_{33}^{(2)}$ of eq.~(\ref{33braket_two_loops}). This relation stems from the fact that the action of $\hat{H}_{3\to3}$ is symmetrised under permutations of the three Reggeons, as both the target and projectile wavefunctions are symmetric. 

The new type of contribution at three loops, Fig.~\ref{fig:1to3}, arises from the term involving $\hat{H}_{3\to 1}$ in eq.~(\ref{eq:mhatNNLL}),
\begin{equation}
\braket{j_{1}|\hat{H}_{3\to 1}|i_{3}}\!=\!\frac{1}{36}\left(-\frac{1}{\eps^3} + {70f_\eps}+{\cal O}(\epsilon^2) \right)
d_i
\braket{j_{1}|i_{1}},\label{eq:h133loops}
\end{equation}
and a similar one involving $\hat{H}_{1\to 3}$, obtained upon replacing 
$d_i\equiv \frac{d_{AR_i}}{N_{R_i}}\frac{1}{C_{R_i}}$ in (\ref{eq:h133loops}) by $d_j$.
Given the representation $R_i$ we defined respectively, $N_{R_i}$, $C_{R_i}$ and $d_{AR_i}$ as the dimension,
the quadratic, and the quartic Casimirs, 
\begin{align*}d_{AR_i}&\equiv \frac{1}{6}\sum_{\sigma\in\mathcal{S}_3}\text{tr}\left(F^aF^bF^cF^d\right)\text{tr}\Big(\mathbf{T}^a\mathbf{T}^{\sigma{(b)}}\mathbf{T}^{\sigma{(c)}}\mathbf{T}^{\sigma{(d)}}\Big)\end{align*}
with $\mathbf{T}^a$ and $F^a$, respectively,  the generators in the $R_i$ and the adjoint representations. 
In contrast to~(\ref{33braket_two_loops})-(\ref{eq:h333loops}), eq.~(\ref{eq:h133loops}) does not involve a matrix in colour space, but is simply proportional to $\braket{j_1|i_1}$. This is due to the fact that colour is carried by a single Reggeon on either the target or projectile sides. This property holds for all terms in the second and third lines of~(\ref{eq:mhatNNLL}) at any order \footnote{We thank Simon Caron-Huot for this important observation}.

\begin{figure}
  \subfloat[\label{fig:H1331colour}]{\includegraphics[width=0.12\textwidth]{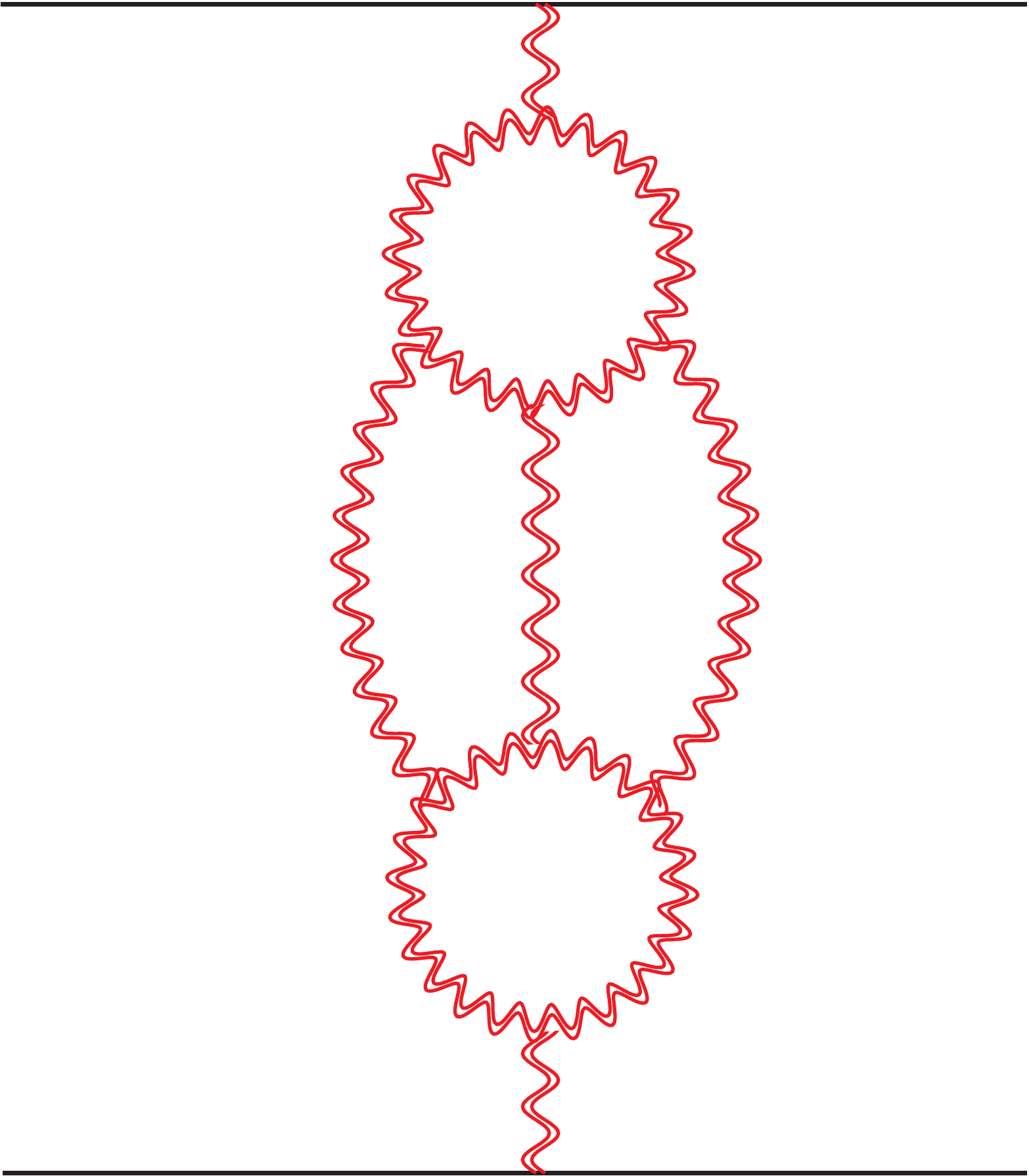}}\,\hspace{10pt}
  \subfloat[\label{fig:H13colour1}]{\includegraphics[width=0.12\textwidth]{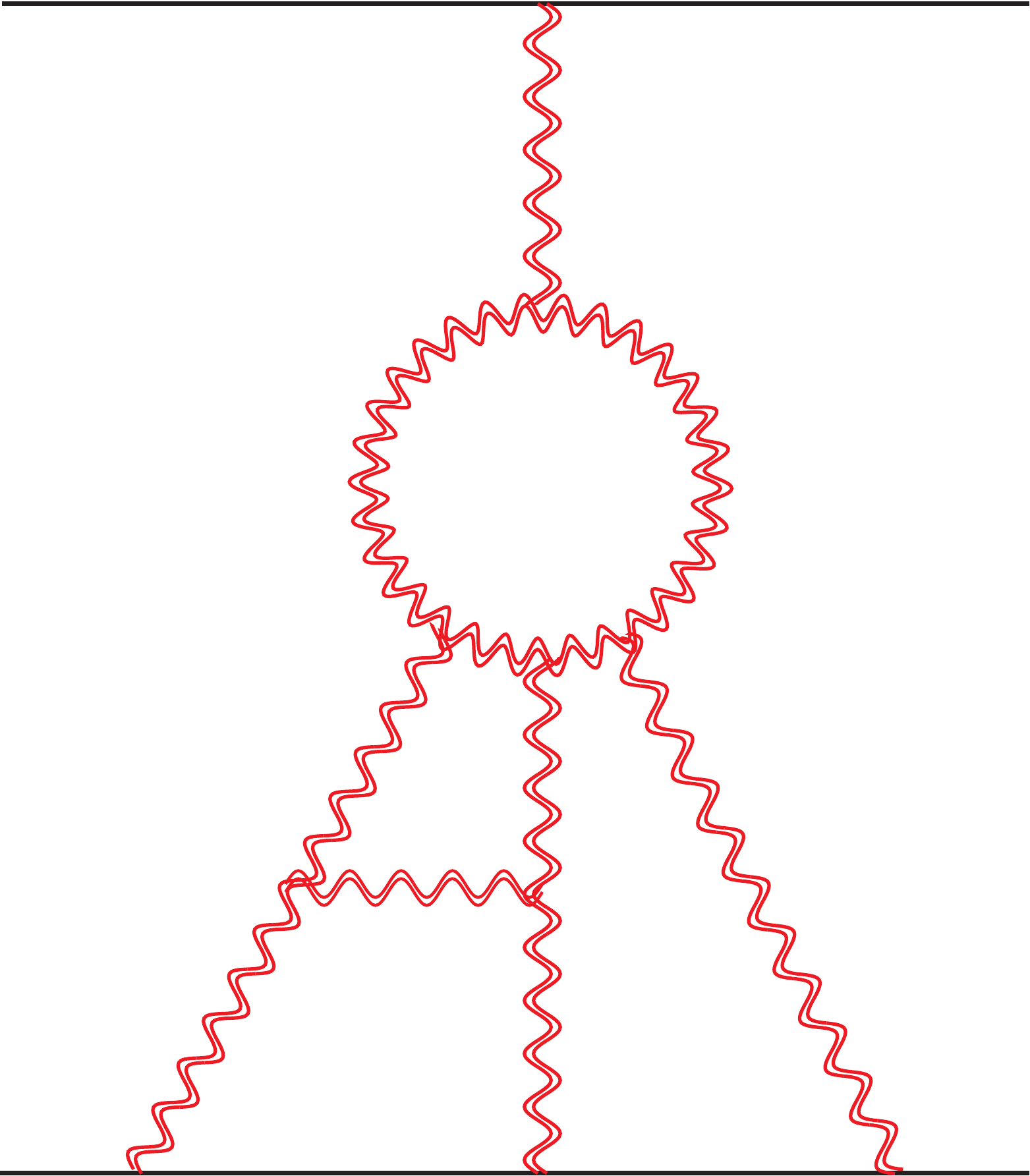}}\,\hspace{10pt}
  \subfloat[\label{fig:H13colour2}]{\includegraphics[width=0.12\textwidth]{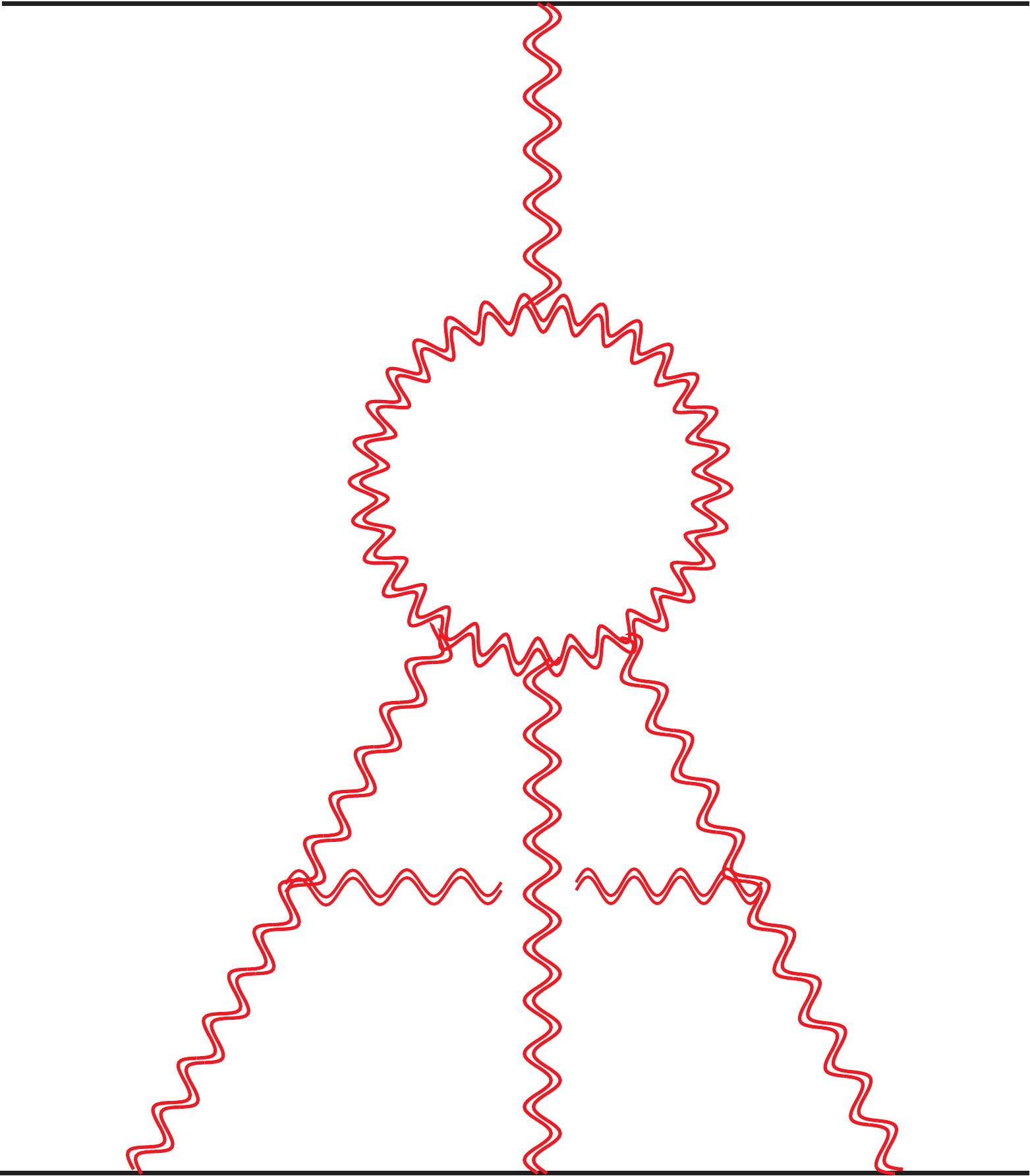}}\,
\caption{Four-loop mixing between a single and three Reggeons states through $\hat{H}_{1\to 3}$ and/or $\hat{H}_{3\to 1}$.}
\end{figure}

{\it \textbf{Four loops.}} According to~(\ref{eq:mhatNNLL}),  the four-loop NNLL amplitude $\hat{\mm}^{(-,4,2)}$ is a sum of four terms:
\begin{eqnarray}
  && \frac{i}{2s}\hat{\mm}^{(-,4,2)}\!=\frac{r_\Gamma^4\pi^2}{2}\bigg[
\braket{j_{1}|\hat{H}_{3\to 1}\hat{H}_{1\to 3}|i_{1}}
+\braket{j_{3}|\hat{H}_{3\to 3}^{2}|i_{3}}
\nonumber\\
  &&\qquad\,\,\,  +\braket{j_{1}|\hat{H}_{3\to 1}\hat{H}_{3\to 3}|i_{3}}+\braket{j_{3}|\hat{H}_{3\to 3}\hat{H}_{1\to 3}|i_{1}}\bigg].\label{eq:oddNNLLDef}
\end{eqnarray}
Their calculation is the main result of this paper. Despite the high loop order, the integration in transverse-momentum space is relatively straightforward: all integrals depend on a single scale, $t$, and involve up to four propagators in a loop; they can be performed~\cite{upcoming} using known techniques~\cite{Lee:2013mka,*Ruijl:2017cxj}. In turn, simplifying the colour structure of eq.~(\ref{eq:oddNNLLDef}) is the main difficulty~\cite{upcoming}. The first term, Fig.~\ref{fig:H1331colour}, yields
\begin{eqnarray}\label{3113}
&&\braket{j_{1}|\hat{H}_{3\to 1}\hat{H}_{1\to 3}|i_{1}}=\frac{1}{432}\bigg[ -\left(\frac{C_A^4}{12}+\frac{d_{AA}}{N_A}\right)\frac{1}{\eps^4}\\& 
&\hspace*{50pt}+\left(\frac{101}{6}C_A^4+220\frac{d_{AA}}{N_A}\right)\frac{f_\eps}{\eps}
+{\cal O}(\eps)
\bigg]\braket{j_{1}|i_{1}}\,,\nonumber
\end{eqnarray}
involving the quartic Casimir in the adjoint representation, $d_{AA}$.

\begin{figure}
  \subfloat[]{\includegraphics[width=0.12\textwidth]{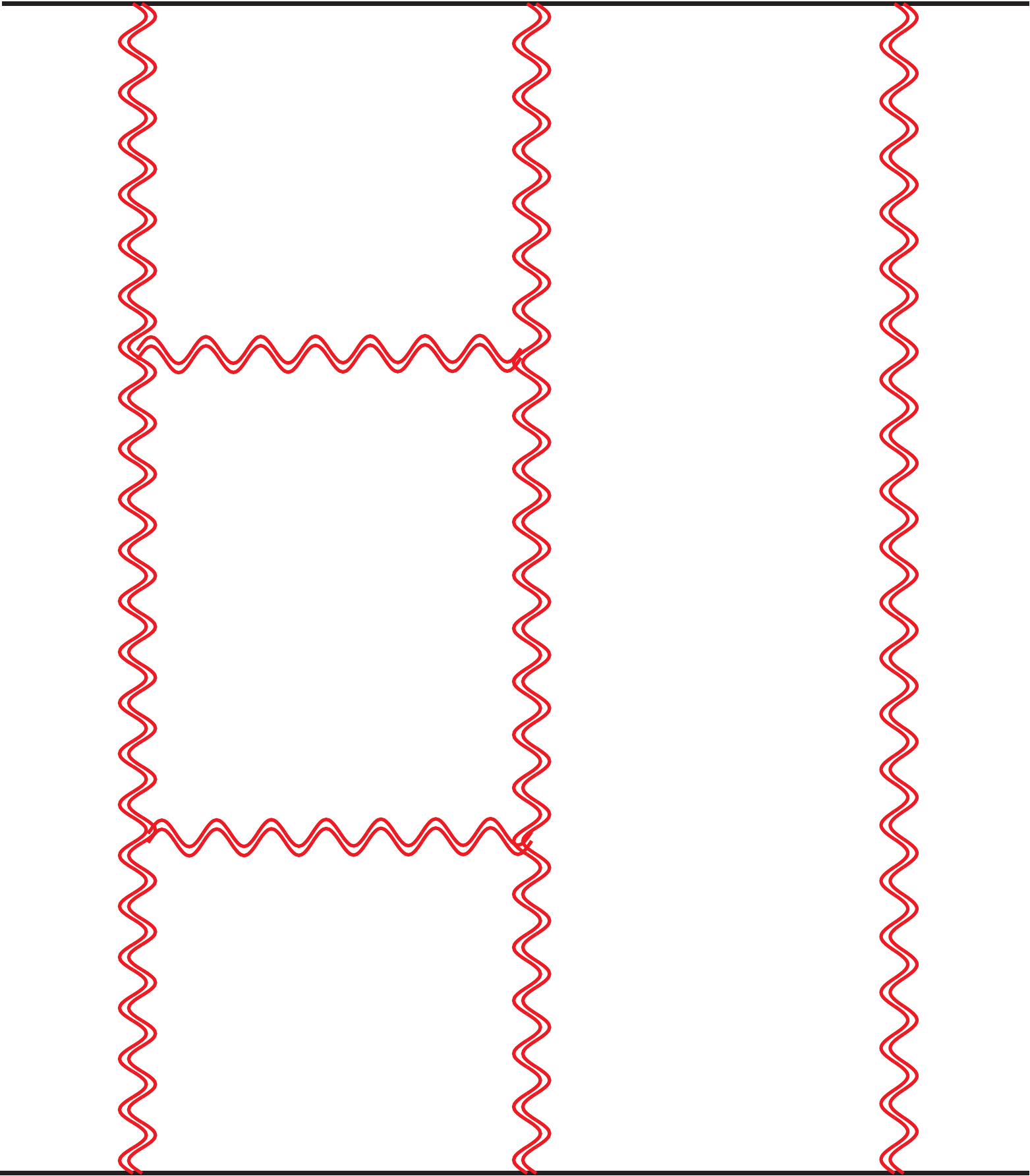}}\,\hspace{15pt}
  \subfloat[]{\includegraphics[width=0.12\textwidth]{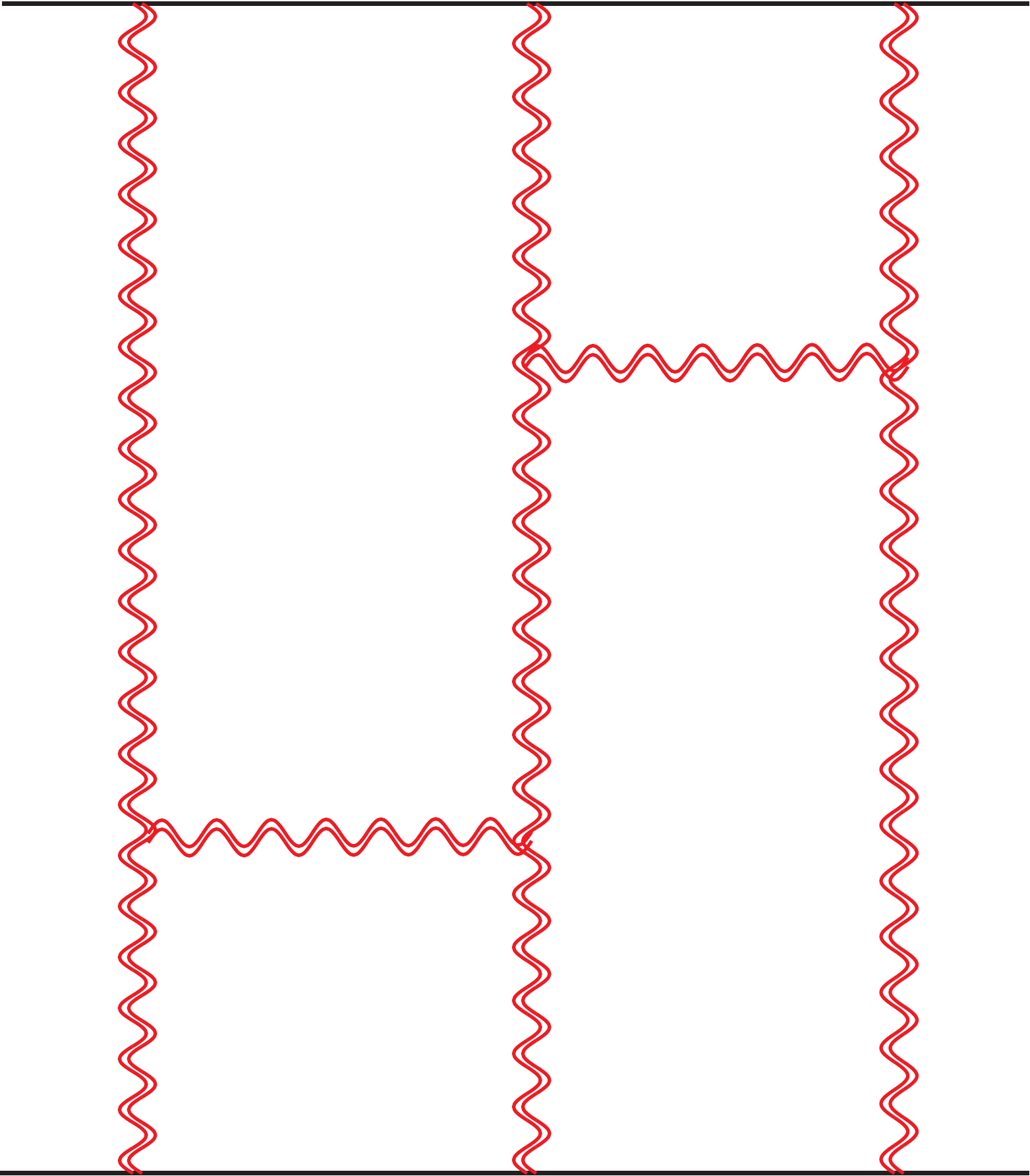}}\,
\caption{Four-loop evolution of the three Reggeon state with two insertions of the $\hat{H}_{3\to 3}$ Hamiltonian.\label{fig:H33colours}}
\end{figure}

The second term in~(\ref{eq:oddNNLLDef}) contains a repeated application of the $\hat{H}_{3\to 3}$ Hamiltonian, yielding two independent colour structures, Fig.~\ref{fig:H33colours}, each multiplying a different integral.  
This breaks the permutation symmetry of the three-Reggeon wavefunction which was responsible for the simple structure in~(\ref{eq:h333loops}), where $\tsu$ entered solely via $\mathbf{C}_{33}^{(2)}$.  The resulting colour structure is thus more complex
\begin{eqnarray}
 && \braket{j_{3}|\hat{H}_{3\to 3}^2|i_{3}}\!=\!\frac{1}{144}\bigg[\frac{\mathbf{C}_{33}^{(4,-4)}}{\eps^4} + \frac{2f_\eps}{\eps}\mathbf{C}_{33}^{(4,-1)} +{\cal O}(\eps)\bigg]\braket{j_{1}|i_{1}} \nonumber\\
  &&\hskip3pt\mathbf{C}_{33}^{(4,-4)} =6\left(17C_A\tts-6C_A^2-6(\tts)^2\right)\mathbf{C}_{33}^{(2)}
\nonumber
\\
&& \hskip11pt-\frac{3}{4}\tsu(\tts)^2\tsu  + \frac{25}{144}C_A^4+\frac{1}{3}\frac{d_{AA}}{N_A}-3C_A(d_i+d_j) \nonumber\\
 &&\hskip3pt \mathbf{C}_{33}^{(4,-1)}= 18\left(521C_A\tts-300C_A^2-220(\tts)^2\right)\mathbf{C}_{33}^{(2)} \nonumber\\  \label{3333}
&&\hskip11pt -101\mathbf{C}_{33}^{(4,-4)}
\end{eqnarray}
involving an additional term, $\tsu(\tts)^2\tsu$.

The third term in~(\ref{eq:oddNNLLDef}) gives rise to two distinct colour structures, Figs.~\ref{fig:H13colour1} and~\ref{fig:H13colour2}, depending on which Reggeons are acted upon by $\hat{H}_{3\to 3}$. As at three loops  in~(\ref{eq:h133loops}), the result is proportional to the tree-level amplitude: specifically, Fig.~\ref{fig:H13colour2} vanishes, while Fig.~\ref{fig:H13colour1} yields
\begin{equation}
\label{3133}
\braket{j_{1}|\hat{H}_{3\to 1}\hat{H}_{3\to 3}|i_{3}}=
\frac{C_Ad_i}{144}\left[\frac{1}{\eps^4} -208\frac{f_\eps}{\eps}
+{\cal O}(\eps)
\right] \braket{j_{1}|i_{1}}\,.
\end{equation}
Finally, the fourth term in~(\ref{eq:oddNNLLDef}) can be simply  obtained from~(\ref{3133}) upon replacing $d_i$ by $d_j$.

Remarkably, $f_\eps$ of~(\ref{33braket_two_loops}) features at three loops in~(\ref{eq:h333loops})-(\ref{eq:h133loops}) and four loops in~(\ref{3113})-(\ref{3133}), all manifesting the simple relation between the coefficients of transcendental weight three and four (this does not extend to weight five).
Adding up the four contributions to~(\ref{eq:oddNNLLDef}), the four-loop NNLL reduced amplitude is found to be
\begin{eqnarray}
\hat{\mm}^{(-,4,2)}\! \!&=&\frac{r_\Gamma^4\pi^2}{144}
\left[
\mathbf{C}_{\mathcal{M}}^{(-4)}\frac{1}{\eps^4}
+\mathbf{C}_{\mathcal{M}}^{(-1)}\frac{f_{\epsilon}}{\eps} +{\cal O}(\eps)\right]\mTree  \nonumber \\ \label{eq:ampResult}
 &&\hspace{-48pt}\mathbf{C}_{\mathcal{M}}^{(-4)}=\frac{1}{2}\mathbf{C}_{33}^{(4,-4)}-\frac{C_A^4}{72}-\frac{1}{6}\frac{d_{AA}}{N_A}+\frac{1}{2}C_A(d_i+d_j) 
\\ \nonumber
 &&\hspace{-48pt}\mathbf{C}_{\mathcal{M}}^{(-1)}= \mathbf{C}_{33}^{(4,-1)}+\frac{101C_A^4}{36}+\frac{110}{3}\frac{d_{AA}}{N_A}-104
C_A(d_i+d_j) \,.
\end{eqnarray}
Interestingly, while the separate contributions have planar components, we find that $\hat{\mm}^{(-,4,2)}$ is non-planar.

{\bf Infrared divergences.}
It has long been recognised that the exponentiation of high-energy logarithms is interlinked with that of infrared singularities~\cite{Korchemskaya:1994qp,*Korchemskaya:1996je,DelDuca:2011ae,Caron-Huot:2013fea,DelDuca:2013ara,DelDuca:2014cya}
and the interplay between the two has been instrumental~\cite{Caron-Huot:2017fxr,Almelid:2017qju,Caron-Huot:2017zfo,Caron-Huot:2020grv,Gardi:2019pmk}. We now take another step in this direction: comparing the NNLL amplitude with the known infrared structure provides an independent check of all multiple poles in $\epsilon$, while the single pole allows us to extract the four-loop soft anomalous dimension at NNLL.

It is well known that infrared divergences in amplitudes factorise and exponentiate~\cite{Catani:1998bh,*Sterman:2002qn,*Aybat:2006wq,*Aybat:2006mz,Gardi:2009qi,*Gardi:2009zv,*Becher:2009cu} according to   
\begin{equation}
	\mathcal{M} = \mathbf{Z}\,\cdot \mathcal{H}, \qquad 
\mathbf{Z} =  \mathcal{P}\exp\left\{-\frac{1}{2}\int_0^{\mu^2}\frac{d\lambda^2}{\lambda^2}\mathbf{\Gamma}\right\}\,,
\label{eq:IRfactor}
\end{equation}
where $\mathcal{H}$ is an infrared-renormalized hard amplitude, which is finite, and  $\mathbf{\Gamma}$ is the soft anomalous dimension. In the high-energy limit the latter takes the form~\cite{DelDuca:2011ae}
\begin{equation}\label{softADdef}
\mathbf{\Gamma} =\frac{\gamma_K}{2}\left[L\tts + i\pi\tsu\right] +  \Gamma_i + \Gamma_j+\mathbf{\Delta},
\end{equation}
where $\Gamma_i$ is defined in~(\ref{Zi}) and $\mathbf{\Delta}$ represents non-dipole corrections starting at three loops~\cite{Gardi:2009qi,*Gardi:2009zv,*Becher:2009cu,Almelid:2015jia}.
The dipole contribution is well known \cite{DelDuca:2011ae,DelDuca:2014cya}: the $L\tts$ term contributes starting at LL, while the $i\pi\tsu$ and $\Gamma_{i}$ terms start at NLL. Here we focus on $\mathbf{\Delta}$, which we expand as
\begin{equation}
	\mathbf{\Delta} = \sum_{n=3}^\infty  \left(\frac{\al_s}{\pi}\right)^n\sum_{m=0}^{n-1}L^m
	\mathbf{\Delta}^{(n,m)}\,.\label{eq:expansionDelta}
\end{equation}
The NLL tower $\mathbf{\Delta}^{(n,n-1)}$ is purely imaginary, and was recently obtained to all orders~\cite{Caron-Huot:2017zfo} from the two-Reggeon amplitude. It starts contributing at four loops~\cite{Caron-Huot:2013fea}, where
\begin{equation}
\mathbf{\Delta}^{(4,3)}= -i\pi\frac{\zeta_3}{24}[\tts,[\tts,\tsu]]\tts.
\end{equation}
The NNLL tower $\mathbf{\Delta}^{(n,n-2)}$ starts at three loops, with a purely imaginary contribution, 
\begin{equation}
 \mathbf{\Delta}^{(3,1)} =i\pi\frac{\zeta_3}{4}[\tts,[\tts,\tsu]],
\end{equation}
determined in \cite{Caron-Huot:2017fxr} on the basis of the 
three-loop calculation of ${\bf \Delta}$ in general kinematics 
\cite{Almelid:2015jia}.
Thanks to the present calculation
of the four-loop amplitude, we obtain the first non-trivial real contribution to this tower,  $\text{Re}[\mathbf{\Delta}^{(4,2)}]$. 
To this end we invert~(\ref{eq:IRfactor}),
and use~(\ref{eq:ampBraket}) and~(\ref{eq:reducedAmpBraket}) to restore the Regge trajectory, obtaining
\begin{equation}
\mathcal{H} = Z_iZ_j\mathbf{Z}^{-1}e^{\al_g\tts L}\hat{\mathcal{M}}.\label{eq:IRinvert}
\end{equation}
The right-hand side of (\ref{eq:IRinvert}) at NNLL depends on $D_i^{(k)}$ in~(\ref{NNLL_single_Reggeon}) and on two- and three-loop Regge-trajectory coefficients, which are all theory-specific.
However, the poles in $\hat{\mathcal{M}}$ are universal \cite{upcoming}, so upon requiring that the hard function $\mathcal{H}^{(4,2)}$ is finite we verify $\mathbf{C}_{\mathcal{M}}^{(-4)}$ in (\ref{eq:ampResult}) and~get
\begin{align}
  \text{Re}\left[\mathbf{\Delta}^{(4,2)}\right]=&\,\,\zeta_2\zeta_3\,\mathbf{C}_{\Delta}^{(4,2)}\,,\label{eq:SADresult}
\end{align}
where the colour structure can be written compactly as
\begin{align}
\mathbf{C}_{\Delta}^{(4,2)}=&\frac{1}{4}\tts[\tts,(\tsu)^2]+\frac{3}{4}[\tsu,\tts]\tts\tsu\nonumber\\&+\frac{d_{AA}}{N_A} -\frac{C_A^4}{24}\,.
\end{align}
As expected, $\text{Re}[\mathbf{\Delta}^{(4,2)}]$ is non-planar.
This result holds for any gauge theory.

Equations (\ref{eq:ampResult}) and  (\ref{eq:IRinvert})  can be further used to determine the finite terms in the hard function at four loops.
In QCD, this cannot yet be done as the three-loop trajectory is unknown. 
In \hbox{$\mathcal{N}=4$} super-Yang-Mills, however, it is known~\cite{Caron-Huot:2017fxr} from the three-loop 4-point amplitude~\cite{Henn:2016jdu}. Thus we can deduce
\begin{align}
  \text{Re}\left[\mathcal{H}^{(4,2)}_{\mathcal{N}=4}\right]
  =&\left[\,\frac{C_A^4}{128}\,\zeta_3^2+\frac{3}{16}\,\zeta_4\zeta_2\,\mathbf{C}_{\Delta}^{(4,2)}\,\right]\,\mTree.\label{eq:hardResult}
\end{align}
The planar limit, $N_c\to\infty$, is already known \cite{Anastasiou:2003kj,*Bern:2005iz,*Drummond:2007aua}; the non-planar correction, the second term, is a new result.

{\bf Conclusions.}
We used an iterated solution of the rapidity evolution equation to derive new results for the Regge limit of $2\to 2$ gauge-theory scattering amplitudes in general colour representations at four loops.
Specifically,  we presented results for the signature odd (real) part of the reduced amplitude in eq.~(\ref{eq:ampResult}) at NNLL accuracy, a result which is valid in full colour for any gauge theory. Interestingly, the result is entirely non-planar. It involves quartic Casimirs associated with the representations of the scattered partons as well as a purely adjoint one.
We observe an interesting relation between the transcendental weight 3 and 4 contributions, already visible at two loops~(\ref{33braket_two_loops}). This relation holds also at three loops, and at four loops it amounts to a relation between the $1/\epsilon$ pole term and the finite terms.

Matching the result to the known exponentiation properties of soft singularities provides a powerful check on the computation, and 
allows us to extract the four-loop NNLL soft anomalous dimension, presented in~(\ref{eq:SADresult}). 
As a by-product we determine the finite terms of the $\mathcal{N}=4$ amplitude, eq.~(\ref{eq:hardResult}).
Regge limit results have proven instrumental in bootstrapping the three-loop soft anomalous dimension in general kinematics~\cite{Almelid:2017qju}. Our result paves the way for a four-loop bootstrap. Ref.~\cite{Becher:2019avh} analysed the colour structure of the soft anomalous dimension, incorporating the recently computed four-loop cusp anomalous dimension \cite{Henn:2019swt,vonManteuffel:2020vjv}, which introduces quartic Casimirs. Our results show that the soft anomalous dimension contains additional quartic Casimirs, beyond those associated with the cusp.

{\bf Acknowledgements.}
We would like to thank Simon Caron-Huot for insightful comments and Niamh Maher for discussions 
regarding the colour structure.
EG and GF are supported by the STFC Consolidated Grant ‘Particle Physics at the Higgs Centre’.
GF is supported by the ERC Starting Grant 715049 ‘QCDforfuture’.
CM's work is supported by the Italian Ministry of University and Research (MIUR),
grant PRIN 20172LNEEZ.  LV is supported 
by Fellini Fellowship for Innovation at INFN, funded by the 
European Union's Horizon 2020 research programme under the 
Marie Sk\l{}odowska-Curie Cofund Action, grant agreement no. 754496.

\bibliography{ReggeRefs}
\bibliographystyle{apsrev4-1}

\end{document}